\documentclass[iop]{emulateapj}
\usepackage{natbib,graphicx,amsthm,ulem,color}

\newcommand{\be}{\begin{equation}}
\newcommand{\ee}{\end{equation}}
\newcommand{\beqn}{\begin{eqnarray}}
\newcommand{\eeqn}{\end{eqnarray}}
\newcommand{\bi}{\begin{itemize}}
\newcommand{\ei}{\end{itemize}}

\def\km{{\rm \, km}}
\def\g{{\rm \, g}}
\def\cm{{\rm \, cm}}

\def\rth{R_{\rm tran}}
\def\rtran{{R_{\rm tran}}}
\def\rst{R_{\rm stir}}

\usepackage{ulem}

\defcitealias{GLS}{GLS}

\begin{document}

\title{After runaway: The Trans-hill stage of  planetesimal growth}
\author{Yoram Lithwick\altaffilmark{1}}
\altaffiltext{1}{Department of Physics and Astronomy, Northwestern University, Evanston, IL 60208 and
Center for Interdisciplinary Exploration and Research in Astrophysics
(CIERA)}

\begin{abstract}

When planetesimals begin to grow by coagulation, they enter an epoch of runaway, during which the biggest bodies grow faster than all the others.  The questions of how runaway ends and what comes next have not  been answered satisfactorily. Here we show that runaway is followed by a `trans-hill stage' that commences once the bodies become trans-hill, i.e. once the Hill velocity of the bodies that dominate viscous stirring matches the random speed of the small bodies they accrete. Subsequently, the small bodies' speed grows in lockstep with the big bodies' sizes, such that the bodies remain in the trans-hill state. Trans-hill growth is crucial for determining the efficiency of growing big bodies, as well as their growth timescale and size spectrum.

We work out the properties of trans-hill growth analytically and confirm these  numerically. Trans-hill growth has two sub-stages. In the earlier one, collisional cooling is irrelevant, in which case the efficiency of forming big bodies remains very low ($~0.1$\% in the Kuiper belt) and their mass spectrum is flat. This explains results from previous coagulation simulations for both the Kuiper belt and the asteroid belt. The second sub-stage commences when small bodies begin to collide with one another. Collisional cooling then controls the evolution, in which case the efficiency of forming big bodies rises, and their size spectrum becomes more top-heavy.

Trans-hill growth terminates in one of two ways, depending on parameters. First, mutual accretion of big bodies can become significant and conglomeration proceeds until half the total mass is converted into big bodies. This mode of growth may  explain the size distributions of minor bodies in the Solar System, and is explored in forthcoming work. Second, if big bodies become separated by their Hill radius, oligarchy commences.  This mode likely precedes the formation of fully-fledged planets.

\end{abstract}

\section{Introduction}

  Our understanding of
how planets form remains inadequate. This hinders  attempts to explain
   the many recent discoveries of extra-solar planets,
  proto-planetary disks and debris disks.  
  Planet formation is often
decomposed into a number of stages.  In the first stage,
``planetesimals'' form out of dust embedded in protoplanetary
disks.  
 How that happens is highly uncertain because it depends on
complicated physics such as how particles stick and how they interact
with a turbulent gas disk \cite[see][for a review]{2010AREPS..38..493C}. 
The initial 
planetesimals are often assumed to be kilometer-sized, but
they could be much smaller or larger
than that \citep[e.g.,][]{Johansen:2007}.
  In the second stage, sometimes called
coagulation, the planetesimals attract one another gravitationally,
 merge and grow.  This stage is more easily understood
than the first, because the dominant physical process is simply gravity
\cite[see][hereafter GLS, for a review]{GLS}.  Nonetheless, past studies have
 given different,
  sometimes even conflicting, views on this process.  What happens
after coagulation depends on local conditions. Just beyond
the snow-line,  coagulation is thought to produce cores that then accrete massive
gaseous atmospheres to form gas giants \citep{1996Icar..124...62P}.  In
the terrestrial zone, coagulation produces dozens of sub-earth-sized
bodies that then undergo a velocity instability once they have
accreted half the planetesimals, leading to an epoch of large-scale
chaos and giant impacts \citep{1998Icar..136..304C,Goldreich:2004b}.
And in the asteroid belt, Kuiper belt, and extra-solar debris disks,
coagulation is thought to have been incomplete, perhaps because the
coagulating planetesimals were excited by exterior planets before
forming planets themselves, or because the initial surface density was very low.

It is the second stage, coagulation, that is the topic of this paper.
Our goal is to build a theory that explains the properties of bodies
that ultimately form, such as their number, size distribution,
formation timescale, and {\it efficiency}, where the efficiency is the
fraction of mass in the original planetesimals that ends up in
big bodies.   This theory can then be compared against
  observations, especially of the asteroid and Kuiper belts, whose
  large bodies are thought to be frozen remnants of the coagulation
  process.

Coagulation is usually studied with numerical simulations.  Starting
from an assumed initial state, the planetesimals begin to merge.  The
merging rate depends on gravitational focusing factors, which are
functions of the relative speeds.
The relative speeds are in turn affected by a variety of two-body
processes such as viscous stirring, dynamical friction, and inelastic
collisions.  While there have been many such simulations, using a
variety of techniques \citep[e.g.,][]{1978Icar...35....1G,
  WetherillStewart,Kokubo:1998, Weidenschilling, KenyonandLuu:1999,
  2001Icar..149..235I, 
  2009Icar..204..558M, ormelicarus, 2011Icar..214..671W,
  2011ApJ...728...68S}, the simulations are complicated,
and hence
it is often difficult to disentangle the various effects in order to understand the results and be confident that they are correct. 
Moreover, the results from different groups do not always agree
\citep[e.g.][]{ 2009Icar..204..558M, 2011Icar..214..671W}.
Compounding the difficulty  are a number of fundamental
uncertainties---such as the unknown initial size and velocity
distributions, how collisional fragmentation occurs,
 and when are
gaps  opened in the circumstellar disk.  A theory for coagulation
would be desirable to guide and interpret simulations, and to
determine how sensitive the results are to assumptions.

Early coagulation simulations led to the discovery of runaway growth,
in which the size distribution quickly develops a tail extending to
extremely large sizes \citep[][]{Safronov1972,1978Icar...35....1G,
  WetherillStewart}.  Runaway occurs because, with
gravitational focusing included, the growth rate of bodies ($d\ln
R/dt$) can be an increasing function of their radius $R$ (see also
Section \ref{sec:run}, below).  This implies that the largest bodies
continue to double to infinite size before smaller ones double a
single time.  The properties of runaway growth have been studied
analytically \citep[e.g.,][]{2000Icar..143...74L,2001Icar..150..314M}.
A variety of theories have been proposed to explain how runaway growth
ends, and what comes next \citep[e.g.,][]{IdaMakino,Kokubo:1998,
  ormelicarus,2011ApJ...728...68S}.   It is often thought
  that  runaway accretion is followed by  self-regulated
  oligarchic growth, during which each big body heats its own food. We show here,
  however, that there is a critical intervening stage of growth, which we call the 
  trans-hill stage.

In the absence of an understanding of the trans-hill stage, previous studies
  \citepalias[e.g.,][]{GLS} could not explain two key results of
  conglomeration simulations: the size spectrum and the formation
  efficiency of large bodies. Here we show that  trans-hill growth
  is a critical stage for determining these, and based on our
  analytical understanding we  provide simple scalings to
  explain results of previous numerical simulations.

  During  trans-hill growth, conglomeration can be 
  categorized as collisionless or collisional, with very different
  outcomes for formation efficiency and size spectrum. In the
  collisionless case small bodies rarely collide, while in the collisional
  one
     the small bodies'
  dispersion is reduced by frequent inelastic collisions.  
  Collisionless growth has been often simulated before, but without
  clear interpretation. Collisional growth, on the other hand, is
  just starting to be explored
  \citep{2011Icar..214..671W,Shannon12b}. Our study here provides a
  unified framework for both regimes.

The structure of this paper is as follows.  In Section \ref{sec:eom}
we present the equations of motion. We re-write these  in
Appendix A in a form suitable for numerical integration or analytic
solution.  In Section \ref{sec:qual} we review runaway growth, and
show that it inexorably transitions into trans-hill growth.  We then derive
qualitatively the properties of trans-hill growth.  In Section
\ref{sec:quant} we present exact 
solutions of trans-hill growth, using numerical integrations as well
as analytical 
 self-similar solutions (derived in Appendix B). 
  Section \ref{sec:after} describes what comes after
trans-hill growth. Section \ref{sec:ass} examines our assumptions
 and delineates their range of validity. Section \ref{sec:disc} discusses
applications of our theory to the asteroid and Kuiper belts, and
compares with a number of earlier papers. Section \ref{sec:sum}
summarizes.

 \cite{Shannon12a,Shannon12b} present
  particle-in-box simulations in both the collisionless and
  collisional regimes, without the restrictive assumptions made in the present paper.  
  The results  there
    confirm and refine those of this study. 
   The collisional paper in particular
  focuses on the formation of the Cold Classical Kuiper belt.  It
  presents a new picture where KBOs form out of a very low mass
  planetesimal disk (the ``minimum mass Kuiper belt''), $100$ times less massive than
     the minimum mass solar nebula.

\section{{ Assumptions and} Equations of Motion}
\label{sec:eom}

We examine first the interactions between two groups of bodies, big
ones and small ones, before proceeding to consider a distribution of
bodies.   This `two-groups' approximation has been described
  by \citetalias{GLS} and we follow their notation. Big bodies have
radius $R$ and surface density $\Sigma$; small bodies have radius $s$,
surface density $\sigma$, and velocity dispersion $u$.  All bodies
  have the same bulk density, with $\rho\sim 1$ g cm$^{-3}$. We
make the following assumptions, and then check  for
self-consistency in Section \ref{sec:ass}:
\begin{enumerate}
\item  The small bodies' random speed $u$ satisfies
\be \alpha^{1/2}<u/v_H<\alpha^{-1/2}\ ,\label{eq:ulim}\ee  where 
$\alpha$ 
is the ratio of all bodies' physical radius to their Hill radius ($R/R_H$) and
$v_H$ is the
big bodies' Hill velocity.
Explicitly,  
\be
\alpha\equiv {R\over R_H}\sim { R_\odot\over a}\ll 1 \ ,
\ee
where  the expression $R_\odot/a$  (the ratio of the Sun's radius
to the semimajor axis)
follows from the fact that the Sun's density 
is comparable to that of solid bodies. 
In the asteroid belt
 $\alpha \sim 2\times 10^{-3}$, while in the Kuiper belt
$\alpha \sim 10^{-4}$.
In addition,
\be
v_H\sim R\sqrt{G\rho\alpha} \ ,
\label{eq:vh}
\ee The small bodies' speed can be sub- or super-hill with respect to
the big bodies ($u<v_H$ or $u>v_H$), subject to the constraints that
$u$ is less than the surface escape speed from the big bodies
($u<\alpha^{-1/2}v_H$) and that small bodies are not accreted in the
``very thin disk'' accretion regime ($u>\alpha^{1/2}v_H$).

\item {The big bodies' random speed $v$ is sub-hill
    ($v<v_H$).}

\item $\Sigma\ll \sigma$, and so $\sigma$ is essentially the total
  surface density, and is treated as a constant  during the
    growth.  More stringently, we assume that big bodies
  grow only by accreting small ones. This is certainly
  true at early times, when there are only a few big
  bodies.

\item The small bodies' size $s$ is a constant parameter.  Small
  bodies do not grow, and they  also do not fragment in
  collisions.

\item More than a single big body dominates viscous stirring at a
  given distance from the star, and hence the big body number
  distribution can be treated as a continuous function.  This
  assumption is violated when oligarchy  commences.

\end{enumerate}

With the above assumptions, big bodies grow by accreting small ones at
the rate \citepalias{GLS}, \be {1\over R}{dR\over dt} =
{\sigma\Omega\over\rho R}\alpha^{-1} \cases{ (v_H/u)^2, & if
  $u>v_H$\cr 
  v_H/u, & if $u<v_H$ 
}
      \label{eq:gr}
\ee
 where $\Omega$ is the orbital angular speed around the Sun.

 For the evolution of $u$, small bodies are damped by inelastic
 collisions amongst themselves, and are viscously stirred by the big
 bodies.  The net rate is \citepalias{GLS}.  \be {1\over u}{du\over
   dt}= -{\sigma\Omega\over\rho s} +{\Sigma\Omega\over\rho
   R}\alpha^{-2} \cases{ (v_H/u)^4, & if $u>v_H$\cr v_H/u, & if
   $u<v_H$ \cr } \ .
 \label{eq:ustir}
 \ee 
  If the first term on the right hand side is important the
   system is collisional, otherwise it is collisionless.

 The above equations for two groups of bodies are easily extended to a
 continuous distribution of big bodies.  We denote the cumulative
 number distribution of big bodies as $N(>R)$.  Because we neglect
 accretion of big bodies, the number of big bodies
 remains constant as they grow.  In other words, $N$ satisfies the
 continuity equation \be {\partial N\over\partial t}+V{\partial
   N\over\partial R} = 0 \ , \label{eq:cont} \ee where $V\equiv dR/dt$
 is given in Equation (\ref{eq:gr}).  In place of Equation
 (\ref{eq:ustir}), we 
 {replace $\Sigma\rightarrow d\Sigma$, then set
 \be
 {d(\Sigma/\sigma)\over d\ln R}=
 R^3 {d(N/\eta)\over d\ln R}
 \label{eq:repl}
 \ee and integrate over $dR$}.  
Here $\eta$ is a constant; we shall not need its value because it
is only the ratio $N/\eta$ that is dynamically significant.

 Equations (\ref{eq:gr})--(\ref{eq:repl}) are the equations of motion.
 We solve them both analytically (Appendix B) and numerically
   (see method in Appendix A).   But before presenting the exact solutions, we
   first use simple analytical arguments to show that runaway growth
   inexorably converges towards `trans-hill growth'. This realization
   gives rise to a number of our main results on the mass spectrum and
   formation efficiency (Section \ref{sec:qual}), as  verified by our
   exact solutions (Section \ref{sec:quant}).

\section{From Runaway to Trans-Hill Growth}
\label{sec:qual}

\subsection{Runaway}
\label{sec:run}

We summarize the traditional picture of how runaway growth proceeds
\citep[e.g.,][]{IdaMakino}.  It is typically assumed that bodies of
  some characteristic size (here, $s$) emerge from a dissipating
  protoplanetary gas disk.  The value of $s$ is highly uncertain,
  since it is not even understood how the bodies formed.

  On the timescale that bodies of size $s$
  collide, 
  $ \rho s/(\sigma\Omega)$, they stir each other up to their surface
  escape speed, $u\sim s\sqrt{G\rho}\sim \alpha^{-1/2}v_H\vert_{s}$
  (Equation (\ref{eq:ustir})), where $v_H\vert_s$ is the Hill velocity
  for bodies of size $s$.  On the same timescale, they grow in mass by
  accreting each other.\footnote{It is possible that some bodies grow
    to very large sizes before $u$ is stirred up to the small bodies'
    escape speed, depending on the details of how the small bodies
    formed. We discuss the effect of this on trans-hill growth
    below.} 
  The evolution of those bigger bodies  (radius labelled by $R$) proceeds in the
  runaway regime.  This is because the growth rate when $u>v_H$ is an
  increasing function of $R$, i.e. $d\ln R/dt\propto R$ (Equations
  (\ref{eq:vh}) and (\ref{eq:gr})).  Thus, comparing two big bodies
  with different values of $R$, the bigger one will double in size
  faster than the smaller one, and then will double again even
  faster. It will ultimately reach infinite size---or violate one of
  the assumptions made---before the smaller one has doubled a single
  time. 
  
\subsection{{ From Runaway to} Trans-hill Growth}

Runaway growth requires $u>v_H$.  Since
$v_H$ increases linearly with $R$, when runaway bodies get big enough their Hill velocity
can be sufficiently large that $u<v_H$.  In that sub-hill 
case, big bodies grow in the ``neutral'' regime \citepalias{GLS},
  i.e., since the growth rate is independent of $R$ (Equation
  (\ref{eq:gr})), the distribution function of big bodies $N(>R)$
  maintains its shape while moving to larger $R$.

  In general, some big bodies grow in the runaway regime, and others
  in the neutral regime, depending on their radius: \beqn
  {\rm runaway\  growth,\ if\ }  R<\rth \\
  {\rm neutral\ growth,\ if\ } R>\rth\ , \eeqn where the trans-hill
  radius is \be \rth\equiv
{u\over \sqrt{G\rho\alpha} } \ , \label{eq:rth}
\ee
which is the  radius of the big bodies that have Hill velocity equal to
 the small bodies' speed ($v_H\vert_\rtran=u$).

 As we now show, the big bodies' distribution function is always
 driven to the trans-hill state, defined as $\rst\sim
 \rth$, where $\rst$ is the radius of the big bodies that dominate
 viscous stirring.  Subsequently, the small body velocity dispersion and
   the size of the bodies that dominate stirring grow in lockstep, maintaining
   $\rst\sim\rth$.
    Our argument proceeds
  by considering the two 
   initial situations, $\rst\ll \rth$ or $\rst\gg\rth$, in turn.

\begin{figure}
\centerline{\includegraphics[width=0.5\textwidth]{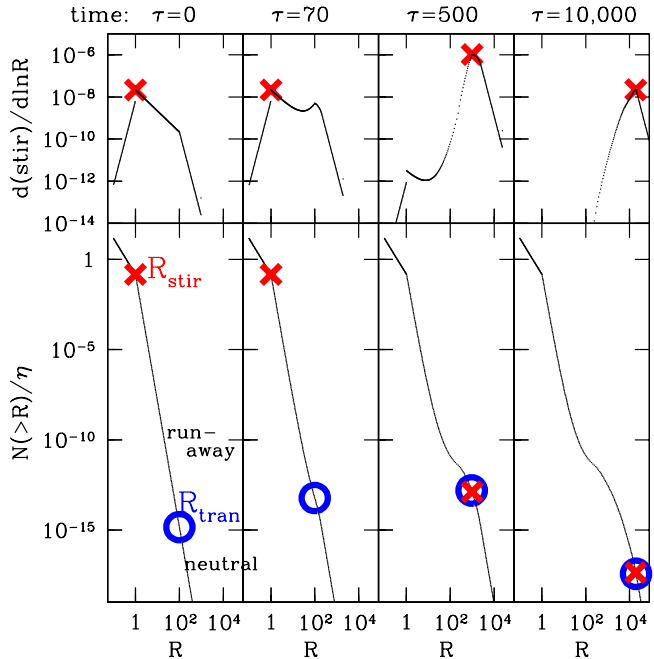}}
\caption{From Runaway to Trans-hill Growth.  Result of a numerical
  integration of the collisionless equations of motion (Equations
  (\ref{eq:gr})--(\ref{eq:repl}) with $s=\infty$ and
  $\alpha=10^{-4}$).  The leftmost panels (top and bottom) show the
  initial conditions, where the bottom panel is the big body
  distribution function, and the top panel is the stirring rate per
  $\ln R$ (the right-hand side of Equation (\ref{eq:ustir}) with
  $\Sigma\rightarrow |d\Sigma/d\ln R|$).  The red cross marks $\rst$,
  defined to be the maximum of the curve in the top panel.  The blue
  circle marks $\rtran\equiv u/\sqrt{G\rho\alpha}$.  
  { Initially $u$ is the escape speed from bodies with $R=1$ (i.e. 
  $\rtran=\alpha^{-1/2}=100$).}
  The panels at
  subsequent times ($\tau=70,500,10^4$,  with $\tau$  the
    scaled time, $\tau\equiv t \sigma\Omega/(\rho\alpha)$; see
    Equation (\ref{eq:taudef})) show that the system evolves from $\rst\ll
\rtran$ to the trans-hill state $\rst\sim \rtran$, and thereafter
remains trans-hill.}
\label{fig:y}
\end{figure}

\bi
\item If initially $\rst\ll \rth$
  (Fig. \ref{fig:y}), as is relevant for the early phases of
  planetesimal growth,  the small bodies are super-hill with respect to the
    stirrers.  All big bodies with $R>\rst$
    repeatedly double in size before the stirring bodies ($\rst$) have
    doubled once.\footnote{
      { Even though very large bodies, with radius above $\rtran$,
        grow in the neutral regime, their doubling rate also exceeds
        that of the stirring bodies.}}  This runaway produces new
    bodies that contribute more to stirring ($\propto \Sigma R^3$,
    Equation (\ref{eq:ustir})) than the original stirrers and these
    new bodies come to dominate the stirring.  Although $\rtran$ also
    grows due to viscous stirring, $\rst$ grows faster, as we
    presently show.  The growth of $\rst$ is at least as fast as the
    size doubling rate at $\rst$, $d\ln \rst/dt \geq d\ln
    R/dt|_{R=\rst}$.  Meanwhile, the heating of small bodies increases
    $\rth$ as $d\ln \rth/dt\sim d\ln u/dt$ (Equation
    (\ref{eq:ustir})).
  We first
  focus on the collisionless case,   inserting $s=\infty$
  into Equation (\ref{eq:ustir}) to get
  \beqn
   {{d\ln \rst/dt}\over{d\ln \rth/dt}} \geq
    \left({{\alpha \sigma}\over\Sigma_{\rm stir}}\right)\,
    \left({\rth\over\rst}\right)^2\, , \nonumber
\label{eq:chase}
\eeqn
where $\Sigma_{\rm stir}$ is the mass density at $\rst$.  If the
efficiency of formation is limited by $\Sigma_{\rm stir}/\sigma \sim
\alpha$, as we show below to be true for the collisionless case, with
time $\rst$ catches up to $\rth$.  { An alternative argument
showing that $\rst$ grows faster
 is that if there were no bodies
  bigger than the original stirring bodies, then, as long as the small bodies are collisionless, 
   the stirring bodies 
  grow at the same rate as $u$, i.e. $d\ln \rst/dt\sim d\ln
  \rtran/dt$.  But by the nature of runaway growth, the bodies bigger
  than the original $\rst$ must grow faster than that.  We conclude
  that, in the collisionless regime,  the ratio $\rst/\rtran$
  approaches unity.}  The simulation in Figure \ref{fig:y} confirms
this.

The same conclusion applies to the collisional case as well, because
collisional damping forces $u$ (and hence $\rtran$) to grow  even
more slowly than in the collisionless case.
Hence $\rst$ catches up with $\rtran$ even more quickly.

\item If initially $\rtran\ll\rst$, the stirring bodies grow in the
  neutral regime.  
  Focusing first on the collisional case, Equation (\ref{eq:ustir})
  with $u \leq v_H$ implies that $\Sigma_{\rm stir}\propto
  u\propto\rth$.  Because the stirring bodies grow neutrally, the
  shape of the distribution function near $\rst$ does not change as
  $\rst$ grows, and hence the number of stirring bodies does not
  change, i.e., $\Sigma_{\rm stir}\propto\rst^3$.  As a result,
  $\rst/\rth\propto 1/\rst^2$, which decreases with time towards
  unity.  The same conclusion applies to the collisionless regime,
  because in that case $u$, and hence $\rth$, will increase even
  faster.  \ei

\subsection{How Trans-hill Growth Proceeds}

{ We have shown that during runaway the big bodies' distribution}
 is driven towards the
trans-hill state \be \rst\sim \rtran \label{eq:th} \ , \ee or
equivalently
\be 
u \, {\sim \,
  {v_H\big\vert_{\rst}}} \ ,
\ee 
{and that after that}
the big bodies continue
to grow in this state.  We call this new stage of growth {\it
  trans-hill growth}.

During trans-hill growth, all bodies
with $R<\rst$ do not grow 
  significantly because those with $R>\rst$ double faster than they
do, i.e.  the distribution becomes frozen in time at small $R$.
Furthermore, bodies with $R>\rst$ follow neutral growth, and their
distribution maintains it shape.  Throughout, $\rth\sim\rst$, and both increase
with time, following {(Equation \ref{eq:gr})} \be \rst \sim
{\sigma\Omega\over\rho\alpha}t \ .
\ee 

We may also evaluate the mass fraction in bodies that dominate
stirring, $\Sigma_{\rm stir}/\sigma$, as a function of $\rst$.
We find,  dropping the ``stir'' subscripts,
\beqn
{\Sigma\over\sigma}\sim \cases{ \alpha, & collisionless ($R<s/\alpha$)
  \cr \alpha^2{R/ s}\, , & collisional ($R>s/\alpha$) \, ,
  \label{eq:spec}
} 
\eeqn 
where the collisionless expression above follows from setting
$d\ln u/dt\sim d\ln \rst/dt$ and dropping the collisional term {in
  Equation (\ref{eq:ustir})}; and the collisional expression follows
from dropping the left-hand side of Equation
(\ref{eq:ustir}).\footnote{Because the spectrum is frozen at small
  $R$, Equation (\ref{eq:spec}) gives not only the temporal evolution
  of $\Sigma_{\rm stir}$, but also the frozen spectrum at
  $R<\rst$---that is why we drop the ``stir'' subscript. } The
transition from collisionless to collisional trans-hill evolution
occurs once $\rst$ exceeds a critical size 
\be
R_{\rm stir}\gtrsim    R_{\rm stir}^{\rm
    (col)} = s/\alpha \ \ \ , \label{eq:ttt} 
    \ee 
     because
beyond this size, small bodies collisionally cool faster than the
growth of trans-hill bodies. This transition occurs at $t \sim {\rho
  s/(\sigma\Omega)}$, i.e., the small bodies' collision time.

Following conventional practice, we  define the
power-law index $q$ of the differential number distribution
 \beqn {dN\over dR}&\propto& R^{-q} \\
&\propto& \Sigma(>R)R^{-4} \nonumber \ , 
\eeqn
in which case we have
 \be q=\cases{ 4, &
  collisionless ($R<s/\alpha$)\cr 3, & collisional ($R>s/\alpha$)
\label{eq:q}
}
\ee

\section{Exact Solutions}
\label{sec:quant}
 
We present exact solutions of the equations of motion.
We first summarize the analytical self-similar solutions  to these
  equations (derived in Appendix B), and then integrate the equations
numerically for a number of cases---both when the initial conditions
are as given by the self-similar solution, and for more realistic
initial conditions.
 
 \subsection{Analytical self-similar solutions}
 \label{eq:ass}
 It is shown in the appendices that the equations of motion { (Equations
   (\ref{eq:gr})--(\ref{eq:repl}), or equivalently Equations
   (\ref{eq:taudef})--(\ref{eq:contapp}))} admit self-similar
 solutions when the radii $R$ are scaled relative to $\rth$, and when
 the stirring is either in the collisionless or collisional regime.
 In either case, there is a single free parameter, { the power-law
   exponent for bodies that grow neutrally, i.e., the value of
   $\gamma$ such that $N\propto R^{-\gamma}$ at $R>\rth$.}  {The
   value of $\gamma$ remains frozen in time by the nature of neutral
   growth, as long as the assumptions of Section \ref{sec:eom} remain
   valid.  However, the value of $\gamma$ in real disks is difficult
   to ascertain from first principles, and likely depends both on how
   the bodies form and on the early stages of coagulation}.
 Fortunately, the solutions are quite insensitive to $\gamma$ as long
 as $\gamma\gg 3$, {or equivalently as long as} stirring is
 dominated by bodies at finite sizes, not by those with $R = \infty$.

 The exact self-similar solutions { (displayed in Equations
   (\ref{eq:ss1}) and (\ref{eq:ncol}) and graphed in Figures
   \ref{fig:ncss} and \ref{fig:colss}) }
confirm what was derived qualitatively in
Section \ref{sec:qual}.  In particular, from the functional dependence
of $N$ on $R$, one sees that stirring is maximized near $R\sim
\rtran$, and hence $\rst\sim\rtran$.  In addition, the differential
mass fraction at $R\ll\rtran$ is  consistent with Equation (\ref{eq:spec}):
\beqn {d\over d\ln R}{\Sigma\over\sigma}=-3g_{\rm nocol}(\gamma)\alpha , {\rm
  \ collisionless} 
\label{eq:nocolsig} \\
{d\over d\ln R}{\Sigma\over\sigma}=-2g_{\rm col}(\gamma)\alpha^2{R\over s}  \ , {\rm \ collisional} \ .
\label{eq:colsig}
\eeqn The { order-unity }functions $g_{\rm nocol}(\gamma)$ and
$g_{\rm col}(\gamma)$ are defined in Appendix B.  They asymptote to
$g_{\rm nocol}(\gamma\rightarrow\infty)=g_{\rm
  col}(\gamma\rightarrow\infty)=10$.

\subsection{Numerical solutions}
\label{eq:ns}

\begin{figure}
\centerline{\includegraphics[width=0.5\textwidth]{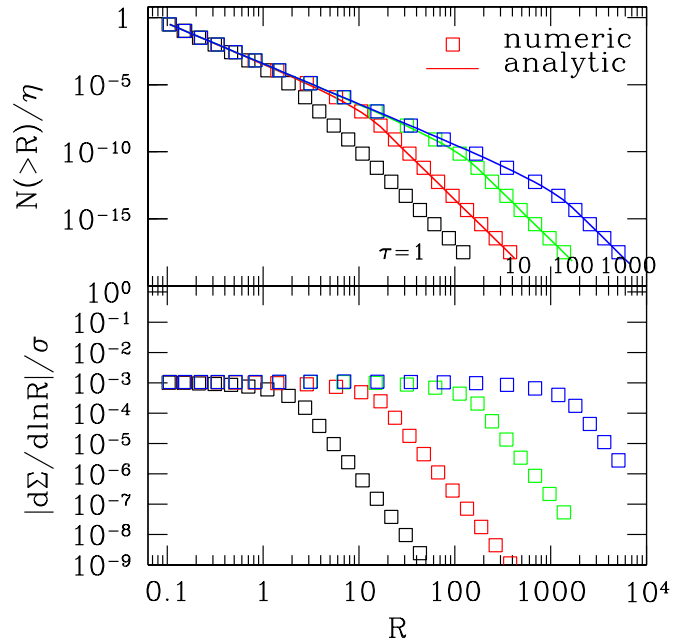}}
\caption{
{Collisionless
    evolution  of Equations (\ref{eq:gr})--(\ref{eq:repl}), 
     initialized with  the self-similar
        size distribution  at scaled time $\tau\equiv
         t\sigma\Omega /(\rho\alpha)=1$; $R$ is measured in units
         of the initial value of $\rtran$.
    {\it Top panel:} The
  cumulative number distribution of big bodies (normalized by 
  constant $\eta$) is plotted at three subsequent  times.
   The numerical and analytic self-similar solutions agree.
   The
    initial parameters are $s=\infty$, $\alpha=10^{-4}$, and
  $\gamma=7$, i.e., $N\propto R^{-7}$ at large $R$.  {\it Bottom
    panel:} The differential
  mass distribution, derived from the top panel via
   Equation
  (\ref{eq:repl}).
  }
    {
   At each time, the radius of bodies that dominate stirring ($\rst$) is  near the break in  
    $N$ (or $\Sigma$) from one power-law slope to the other, 
    and the small body speed is such that  $u$ is trans-hill relative
    to those stirring bodies (i.e., $\rtran/\rst=$  order-unity constant).
    Collisionless evolution leads to a size
    spectrum $dN/dR \propto R^{-4}$ at sizes below $\rst$, and the
    efficiency of forming large bodies is low, $d(\Sigma/\sigma)/d\ln R\sim 10\alpha$.
    } 
   }
\label{fig:ncss}
\end{figure}
\begin{figure}
\centerline{\includegraphics[width=0.5\textwidth]{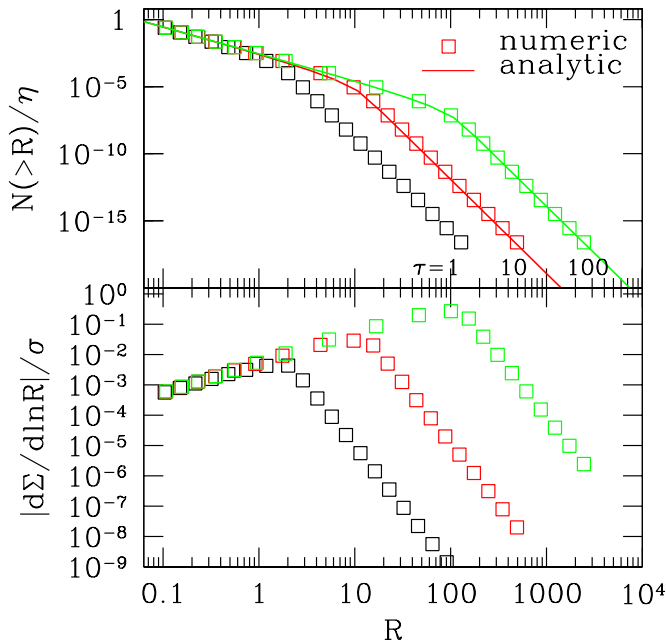}}
\caption{Collisional 
evolution:   Similar to Figure \ref{fig:ncss}, but showing the collisional 
 case with parameters  $s=0.1\alpha$ ({in units where the initial $\rtran=1$})  and, as before, 
$\alpha=10^{-4}$ and $\gamma=7$.   The size spectrum
is $dN/dR \propto R^{-3}$ at sizes below $\rst$, and the efficiency of
forming large bodies is much higher than the collisionless case.
}
\label{fig:colss}
\end{figure}

\begin{figure}
\centerline{\includegraphics[width=0.5\textwidth]{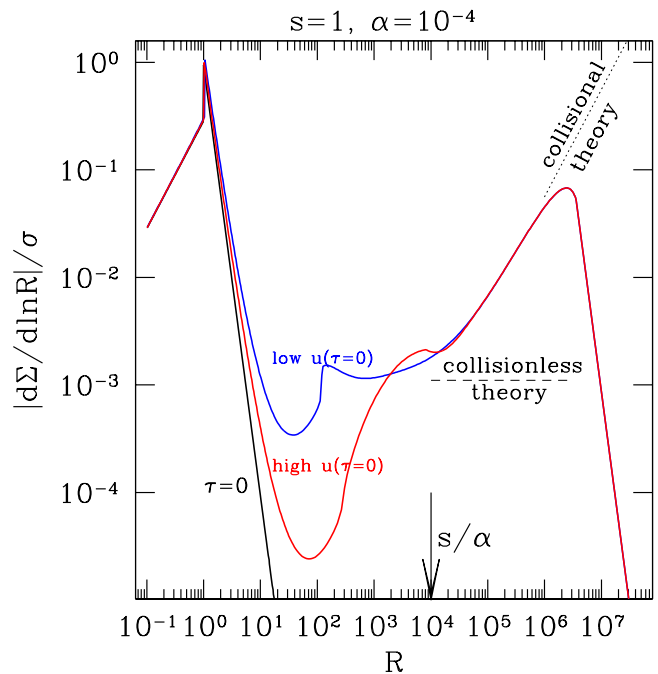}}
\caption{{ From runaway to trans-hill}.  Two integrations were initialized with  nearly all
    bodies at the same size, $R=1$  (the black profile marked 
   $\tau=0$).
  Numerical integrations of  the equations of motion
     are plotted at scaled time $\tau=2.5\times 10^6$.  For the
  red (high $u$) curve, the initial $u$ is the surface escape speed
  from bodies with $R=1$ (i.e., $\rtran=1/\sqrt{\alpha}=100$); for the
  blue curve, it is ten times smaller ($\rtran=10$).  Both
  integrations
  {begin in the runaway stage, and converge}
  onto
    the collisional self-similar solution
    (Equation (\ref{eq:colsig}), dotted line)
  when $R\gtrsim s/\alpha$ (Equation (\ref{eq:spec})).
  The low $u$ case follows the collisionless solution 
    (Equation (\ref{eq:nocolsig}), dashed line) for some time
    before collisional cooling becomes important. This explains the
    flat mass spectrum from $R=10^2 - 10^4$.  The high $u$ case
  takes longer to reach the self-similar solution because its initial
  conditions differ more from the trans-hill state; in fact, it
  effectively skips the collisionless solution entirely. 
    }
\label{fig:converge}
\end{figure}

Figure \ref{fig:ncss} shows a numerical integration of the equations
of motion in the
collisionless case ({$s=\infty$}),  initialized  with the collisionless self-similar
  solution (Equation \ref{eq:ss1}) at scaled time 
 $\tau= 1$.
 It is apparent from the figure that
the evolution remains self-similar at all times, and agrees with the
analytic expression (Equation (\ref{eq:ss1})).  The
  mass distribution per logarithmic bin is indeed constant, 
  in agreement with 
    Equation (\ref{eq:q}).

    Figure \ref{fig:colss} shows an integration for the collisional
    case. It is initialized with Equation (\ref{eq:ncol}) at $\tau=1$,
    has $s=0.1\alpha\times\rtran\vert_{\tau=1}$,
   and other parameters as before. 
The numerical integration agrees with the analytic expression,  and
  shows that the mass distribution is top-heavy with $dN/dR \propto
  R^{-3}$ (Equation (\ref{eq:q})).
  
  In Figure \ref{fig:converge}, we experiment with a  more 
      commonly adopted initial spectrum---one that is
  {more strongly} peaked at a single size ($R=1$).  We perform two
  integrations, both of which have $\alpha=10^{-4}$, $s=1$ ({i.e.,
    $s$ is the same as the location of the peak in the initial size
    distribution}), {and initially $N\propto R^{-7}$ at $R>1$.  The
    two integrations} differ in the initial $u$: one has initial
  velocity equal to the escape speed of bodies with $R=1$, and the
  other has $u$ ten times lower.  Over time, both converge to the
  collisional self-similar solution.  The low $u$ case undergoes a
  temporary phase of collisionless trans-hill growth, when the big
  bodies have yet to grow to $R \geq s/\alpha$.  By contrast, the high
  $u$ case takes longer to reach the trans-hill solution because its
  initial condition is more discrepant from trans-hill.  In fact, this
  case completely skips the collisionless trans-hill regime.
  {We have also experimented with different values for the initial
    power-law slope of $N$.  The main resulting difference is the
    value of $R$ at which the collisionless trans-hill solution
    commences.  For example, when $N\propto R^{-5}$ at $R>1$ (rather
    that $R^{-7}$ as in Figure \ref{fig:converge}), then the high $u$
    solution follows the collisionless theory from
    $R=10^{2.5}$--$10^4$.}

In summary,  as long as the evolution is collisionless, the
efficiency of forming big bodies is small, with $\Sigma/\sigma\simeq
10\alpha$, independent of $R$. But once the evolution becomes
collisional ({$\rst>s/\alpha$}), the efficiency grows towards unity
as $\rst$ increases.
  
\section{After Trans-hill Growth}
\label{sec:after}

\subsection{ Growth by accreting big bodies: Equal accretion}
\label{subsec:bigbig}

One of the ways trans-hill growth can end is when
accretion of big bodies becomes important, violating assumption  3 in
Section \ref{sec:eom}.
To  establish when that occurs, we calculate the accretion rate of
big bodies, which requires knowing the big
bodies' random speed $v$. Balancing  dynamical friction damping
  due to small bodies  with
   viscous stirring due to other big bodies yields 
   \citepalias{GLS}
   \be {1\over v}{dv\over dt}\sim
{\Omega\over\rho R}\alpha^{-2}\left(-\sigma+\Sigma {v_H\over v}
\right) \sim 0 \ , 
\label{eq:vset}
\ee 
Therefore
 ${v/ v_H}\sim {\Sigma/\sigma}
  \ll 1 $.
  The growth rate by accreting big bodies with $v<v_H$
  is (\citealp{Rafikov2003b},GLS) \be {1\over R} {d
    R\over dt}\big\vert_{\rm big}\sim {\Sigma\Omega\over\rho
    R}\alpha^{-3/2} \ , \ee which is larger than the usual (isotropic)
  sub-hill accretion formula (e.g., Equation (\ref{eq:gr})) because
  big bodies lie in a zero-inclination disk.  Comparing  growth by
    accreting big bodies with  that by
  accreting small bodies, 
  \beqn
  {d\ln R/dt\vert_{\rm big}\over d\ln R/dt\vert_{\rm small}}&\sim& {\Sigma\over\sigma}\alpha^{-1/2}\\
  &\sim& \cases{ \alpha^{1/2}, & collisionless \cr \alpha^{3/2}{\rst/
      s}, & collisional }
\label{eq:sch}
\eeqn 
using the trans-hill
expressions.  We conclude that in the
collisionless regime it is always safe to ignore accretion by big
bodies, in disagreement with \cite{2011ApJ...728...68S} (see also
Section \ref{sec:disc}).  However, in the collisional regime, big body
accretion becomes important {once $\rst$ exceeds a critical value
\beqn
R_{\rm stir}\gtrsim R_{\rm stir}^{\rm (b.b.\ accrete)}&=& s \alpha^{-3/2} 
\nonumber
\\
&\approx& 2800 \km \left({s\over{1\km}}\right)
\left({a\over{1 {\rm AU}}}\right)^{3/2}\, .
\label{eq:requal}
\eeqn
}
 This occurs at
time $t\sim \alpha^{-1/2}\rho s/(\sigma\Omega)$, and at that time the
fraction of mass in big bodies (i.e., the efficiency) is
$\Sigma/\sigma\sim \alpha^{1/2} \sim { 7}\% (a/1{\rm AU})^{-1/2}$.
 
As  shown in
 \cite{Shannon12b}, after  collisional trans-hill accretion
ends, an epoch of ``equal accretion'' begins, during which big
bodies grow by accreting comparable mass in big and in small bodies. 
 Equal accretion terminates when half
  of the mass has been converted into big bodies, $\Sigma\sim\sigma$.

  {The scenario outlined above is applicable as long as oligarchy has not yet
  begun.   
    We discuss oligarchy next.}

\subsection{Oligarchic Growth}
\label{subsec:oligarchy}

{As accretion proceeds, the number of stirring bodies decreases, and hence
 neighboring stirrers become increasingly separated.
Eventually, 
   they become so separated that 
   each small body is predominantly stirred by, and accreted onto, a single big body.
 When that happens, assumption
 5 is violated (Section \ref{sec:eom}) and oligarchy commences \citep[][GLS]{Kokubo:1998}.
 }
{In oligarchy, the} nature of growth is
  modified. Instead of a continuous size spectrum, the largest body
  {in each radial annulus separates from the size spectrum of smaller bodies.}
     The velocity dispersion $u$ is no
  longer trans-hill. 

The value of $\rst$ at which oligarchy starts depends on the size
spectrum, for which we now have a simple model. For trans-hill
velocity dispersion, oligarchy begins when the separation between
adjacent big bodies ($\Delta a\sim \rho R^3/(a\Sigma)$) exceeds their
Hill radius ($R_H\sim R/\alpha$){.}  
If we define the oligarchy parameter 
\beqn {\rm OP}&\equiv&
{ \Delta a \over R_H} \nonumber
\\ &=&
{\rho\alpha R^2\over \Sigma a} \ , \label{eq:op} 
\eeqn
 then oligarchy
begins when OP$\sim 1$.  
 Inserting the efficiency of big body
  formation (Equation (\ref{eq:spec})) 
  into the above, we find that  {trans-hill growth transitions to
  oligarchy once $\rst$ exceeds
   \beqn R_{\rm stir}^{(\rm oligarchy)} \approx \cases{ \sqrt{{\sigma
      a}\over{\rho}} \approx 150\km
\cdot      \sigma_{16}^{1/2}
  \left({a\over{1 {\rm
          AU}}}\right)^{1/2}\, , & {\rm collisionless} \cr
  \left({\alpha\over s}\right){{\sigma a}\over\rho}\approx 120\km
\cdot \sigma_{16}
  \left({s\over{1\km}}\right)^{-1}\, , & {\rm collisional}
  \label{eq:opr}
} \eeqn 
where  $\sigma_{\rm 16}\equiv \sigma/(16\g/\cm^2)$.
  For the MMSN density profile, $\sigma_{16}\approx(a/{1 \rm AU})^{-3/2}$ 
  with an enhancement beyond the snow line by a factor of $\sim 5$.
  The collisionless expression above applies as long  
  $R_{\rm stir}^{(\rm oligarchy)}<R_{\rm stir}^{\rm (col)}\approx s/\alpha$;
  otherwise the collisional one applies.
  However, we caution that depending on  parameter values the
  equal accretion stage discussed in Section \ref{subsec:bigbig} may begin
  earlier than oligarchy, which would lead to a different transition
  criterion.}

 Once oligarchy takes hold, neighboring regions evolve
  independently under the stirring of their respective oligarchs, as
   discussed in Sections 9-10 of \citetalias{GLS}. Neighboring oligarchs
  likely converge in size, their battling leading to scattering or
  merging. The evolution depends to a large degree on the fate of the
  small bodies.
{At late times, small bodies collide at such high speeds that they almost
certainly fragment.  As }
 the small bodies grind down each
  other, the evolution can become increasingly collisional. The small
  bodies may cool and the oligarchs may carve gaps around themselves,
  effectively sabotaging their accretion \citep{Rafikov,levison}. We
  defer considerations of these dynamics to future work.

   If the oligarchs eventually reach the isolation mass, they may undergo
orbital instability and experience giant impacts. The timescale to
form  planets of size $R$ is
\begin{equation}
  t_{\rm impact} \approx
  {{\rho R}\over{\Sigma \Omega}} 
  \sim 10^8 \left({R\over{R_\oplus}}\right)
   \sigma_{16}^{-1} \left({a\over{1 {\rm AU}}}\right)^{3/2} {\rm yrs}\, ,
\label{eq:giantimpact}
\end{equation}
which is very long in the outer solar system.
  
{ Our criterion for oligarchy (OP$\sim 1$) differs from that of
 \citet{2010ApJ...714L.103O}.  They stipulate, based on empirical
 evidence from Monte Carlo simulations,
 that oligarchy begins when the stirring rate by one single large body
 equals its growth rate, and that, in turn equals the small body
 collision rate.  Their definition of oligarchy differs from
 ours: their oligarchy begins when the ratio of small body random
 velocity to Hill velocity of the biggest body reaches a minimum.  In
 fact, their oligarchy is similar to (but still somewhat different
 from) the trans-hill phase.  To some extent, this is a matter of how
 one defines oligarchy. But irrespective of definitions, there should
 be a transition from trans-hill to oligarchic behavior when $\Delta
 a\sim R_H$.  }

\section{Examining assumptions}
\label{sec:ass}

We examine the assumptions  made in deriving the properties of
trans-hill growth (Section
\ref{sec:eom}):

\begin{enumerate}
\item The requirement that $u$ satisfies Equation (\ref{eq:ulim}) is
  equivalent to insisting that we only consider big bodies with
  $\alpha^{1/2}<R/\rth<\alpha^{-1/2}$.  { Bodies that violate these
    restrictions grow more slowly than Equation (\ref{eq:gr}) predicts
    \citep{GLS}, but this has little effect on trans-hill growth.  }

\item { The assumption that $v/v_H<1$ is confirmed by Equation
    (\ref{eq:vset}).}

\item { We ignored growth by accretion of big bodies. When that
    assumption is violated, equal accretion begins (Section
    \ref{subsec:bigbig}).  }

\item We assumed that $s$ is constant. In truth, at late times $u$ can
  become sufficiently large that collisions between small bodies
  fragment them. We do not treat this in detail because the physics of
  fragmentation is complex.  We note, however, that since bodies of
  smaller sizes are typically more resistant to fragmentation (for
  $s\ll $ 1 km), the result of fragmentation is likely that $s$ is a
  decreasing function of the stirring bodies'  sizes. This
  will alter the spectrum of trans-hill accretion (Equation
  (\ref{eq:spec})) in that one should replace the $s$ in that equation
  with the function $s(\rst)$.  A similar remark applies to other
  mechanisms for damping $u$, such as gas drag.  Since $s$ only
  appears in the equations of motion through its damping effect on
  $u$, gas drag may also be modelled by  adopting a much smaller
    effective $s$.

\item  We ignored oligarchy. The transition to oligarchic stage
is described in Section \ref{subsec:oligarchy}.

\end{enumerate}

\section{Applications and Comparisons}
\label{sec:disc}

\subsection{The Asteroid Belt}

  Here we
  compare our results  with 
  published coagulation simulations for the asteroid belt.  
 At a distance of $2$ AU, $\alpha \approx
  2.5\times 10^{-3}$, so the transition from collisionless to collisional
  evolution occurs when large bodies grow beyond $s/\alpha\sim 400\km (s/1\km) $.

The uppermost end of the asteroid mass distribution is roughly flat,
i.e. $d\Sigma/d\ln R\sim$ const for $100 {\rm km}\lesssim R\lesssim
300$km.  But for
$R\lesssim 100$km, $\Sigma$ falls off with decreasing $R$
\citep[][taking a constant albedo of
$0.04$]{2002aste.conf...71J,2005Icar..175..111B}.  Both
\cite{2009Icar..204..558M} and \cite{2011Icar..214..671W} attempt to
reproduce the observed distribution, including the 100km ``bump,'' by
running particle-in-a-box coagulation simulations that are initialized
with single-sized small planetesimals.  But the two reach opposite
conclusions.  \citet{2009Icar..204..558M}, with small plantesimal
sizes ranging from $s=0.6$ to $6\km$, fail to produce the 100km bump:
the final distributions they find are roughly flat ($q\sim 4$) for a
range in $R$ that extends an order of magnitude below 100km.  But
Weidenschilling succeeds when using $s\approx 0.1 \km$.  He attributes
the difference to the erroneous neglect of sub-hill accretion by
Morbidelli et al. In the following, we focus on
  explaining Weidenschilling's results.  \footnote{ An additional
  difference between the simulations of \cite{2009Icar..204..558M} and
  \cite{2011Icar..214..671W} is that the former initialize velocities
  to be the Hill velocity of the initial bodies, whereas the latter
  initialize them to be the escape speed.  This might tend to flatten
  out the post-runaway bump in the simulations of Morbidelli et al.,
  similar to how, in Figure \ref{fig:converge}, the blue curve is
  flatter than the red.  
}

{ Weidenschilling's bump at 100km appears to be due to the transition
  from runaway to collisional trans-hill growth.}  These simulations
resemble the red curve in Fig. \ref{fig:converge}{---with a
  transition directly from runaway to collisional
  trans-hill}---because the initial velocity is the escape speed from
seeds. And like the red curve that shows a transition around $R \sim
s/\alpha$, the final size distribution in \citet{2011Icar..214..671W}
also shows a bump around 50km (his Fig. 8), produced when the small
body dispersion goes from super-hill (runaway) to trans-hill.
 The
value of $\Sigma/\sigma$ at the location of his bump, and the slope to
the right of the bump, are all roughly consistent with what we derive
in this paper.

If the  observed 100km bump in the  asteroid belt is indeed produced by the
transition from runaway to trans-hill growth, that would imply that
the initial mass  of small plantesimals, relative to that in
  current big bodies, was at least $\sigma/\Sigma\sim
1/(10\alpha)\sim 50$ times greater  (Equation
(\ref{eq:nocolsig}))---or even greater if some asteroids were
dynamically ejected \citep{2009Icar..204..558M}.  It would also imply
that the initial small bodies had radii $\sim \alpha\times 100
$km$\sim 0.2$km\footnote{
  The  initial small body radii would be less than $0.2$ km by a factor of
  $\sim 5$
   if the MMSN budget of gas was still present when the bump was formed
   \citepalias[e.g.,][]{GLS}.}  One 
 interesting implication of such a scenario is that the vast
majority of these small bodies could not be dynamically ejected,
because if they were that would correspondingly reduce the number of
big bodies too.  Instead, they had to be ground down to dust which was
then eliminated by radiation forces.  Whether this can occur has yet
to be examined carefully.  But we note that by the
time the asteroids grew to $\sim 100$km, the small bodies would have
begun to collide with one another with collision speeds 
$v_H\vert_{R=100{\rm km}}\sim 5$m/sec.  Such speeds might have been
sufficient to grind all the planetesimals and eliminate them,
preventing further growth of asteroids.

\subsection{The Kuiper Belt}

With a value of $\alpha\approx 10^{-4}$ for the Kuiper belt, the
  transition to trans-hill collisional growth occurs at 
  $10^4\km (s/1\km)$. So with the typical choice  $s=1\km$ and the
  largest bodies observed at $1000\km$, there is little wonder that 
  simulations to date have not probed the collisional regime
  \citep[but see][]{Shannon12b}.

\cite{KenyonandLuu:1999} perform particle-in-box simulations for the
formation of Kuiper belt objects.  Starting from bodies of size $s=80$
m, their simulations produce big bodies with a flat mass distribution
($q=4$) for $3 {\rm km}\lesssim R\lesssim 1000 {\rm km}$  (see,
  e.g. their Fig. 8).   This is roughly consistent
with the observed mass distribution in the Kuiper belt at large sizes
\citep[e.g.,][]{FraserBrown}, and hence may be taken as evidence that
the formation of Kuiper belt objects has been solved.   However,
  recent data have cast doubt on such a simple picture
  \citep{gladman12}.

Our theory explains some aspects of Kenyon \&
Luu's simulations.
  The
growth  they witness should entirely be in the collisionless
regime, and the mass spectrum should be flat, as indeed they observe.
However, the efficiency that they find is $\sim 1\%$, whereas we
predict $\sim 10\alpha\sim 0.1\%$.  We have been unable to resolve
this discrepancy.  In  \cite{Shannon12a}, we repeat
their simulations and find an efficiency of $\sim 0.1\%$.  

\citet{2011ApJ...728...68S}
consider collisionless coagulation both 
  analytically and with numerical simulations.  They conclude
that  the flat mass spectrum  obtained in colllisionless
  simulations arises because big bodies grow equally by accreting big and
small bodies (``equal accretion'').  By contrast, we have shown 
that it is due to trans-hill growth.
In fact, Equation (\ref{eq:sch}) shows that
trans-hill growth  in the collisionless regime prevents equal
accretion from occuring.  Nonetheless, the efficiency 
they derive ($\alpha^{3/4}$),
  happens to be close in numerical value to our
  efficiency of $10\alpha$ for Kuiper belt parameters.
  In \citet{Shannon12a}, we confirm with numerical simulations that account for order-unity
  coefficients
  that it is trans-hill growth rather than equal accretion that
  sets the size spectrum in the collisionless regime.

 A troubling concern with  low-efficiency collisionless growth of
  KBOs is that one needs to grind down $99.9\%$ of the small bodies
  to dust and blow it out with radiation pressure.  As we show in an
  upcoming paper, this is almost impossible to do. This concern is
  alleviated if the initial planetesimal size is significantly smaller
  than $1\km$. The formation  efficiency can then be significantly
  boosted by collisional growth \citep{Shannon12b}. 

\subsection{\citet{ormelicarus}}

  \citet{ormelicarus}
  perform a comprehensive study of conglomeration, at semimajor axes
  ranging from 1 AU to $35$AU.  Their
  numerical algorithm allows them to follow the evolution from 
  runaway  through oligarchy. We observe that there is a
  distinct trans-hill phase in their simulations (e.g., Figs. 8 \& 11
  of that paper). With their parameters of $s= 7.5\km$, $\sigma=16
  \g/\cm^2$,
  one expects the transhill evolution (collisionless below $\rst = 1500
  (a/1{\rm AU})\km$) to lead to a characteristic size spectrum of
  $q=4$ and a formation efficiency of $\Sigma/\sigma \sim 0.05
  (a/1{\rm AU})$. These are indeed observed in their
  results.\footnote{Although \cite{ormelicarus} fit their result with a
    power-law slope of $q=5.5$ ($p=-2.5$ in their notation), it appears that the big bodies are better characterized
    by a flat mass distribution of $q=4$ (see, e.g., their Fig. 13).}
  Above $R = 1500\km (a/1{\rm AU})$, the evolution is collisional.
  For some of their simulations (the ones at small $a$), oligarchy
  enters at around the same point.
  So instead of observing a continuous mass spectrum of $q=3$
  (Equation (\ref{eq:q})), they find that a single body that grows to large
  sizes.

\section{summary}
\label{sec:sum}

Runaway growth ends once the stirring bodies become trans-hill
($v_H\vert_{\rm stir}\sim u$, or equivalently ${\rst}\sim
u/\sqrt{G\rho\alpha}$). Afterwards, $\rst$ and $u$ grow in unison, and
the stirring bodies remain trans-hill.  For $\rst<s/\alpha$,
trans-hill growth is collisionless.  The mass spectrum is flat
($q=4$), and the efficiency low, $\sim 10\alpha$.  But for
$\rst>s/\alpha$, the evolution is collisional and the efficiency grows
in proportion to $\rst$, with a size spectrum that has $q=3$.  The
time {to reach the collisional transition} is comparable to the
small bodies' collision time ($\rho s/\sigma\Omega$).  Our numerical
simulations and self-similar calculations confirm these results, {
  subject to the assumptions made in Section \ref{sec:eom}}.  The
simulations also show that where the {collisionless} trans-hill
spectrum begins depends on the {unknown} initial {
  conditions---especially the initial size spectrum and $u$.  But the
  collisional trans-hill spectrum invariably begins once
  $\rst>s/\alpha$.}

 Having described trans-hill growth, we briefly discussed what
  comes next, when one or more of the assumptions in Section
  \ref{sec:eom} breaks.  Collisional trans-hill growth ends once
$\rst\gtrsim s\alpha^{-3/2}$ (efficiency $\Sigma/\sigma\gtrsim
\sqrt{\alpha}$), because once that happens mutual accretion of big
bodies becomes important (Section \ref{sec:after}).  As we show in a
forthcoming paper \citep{Shannon12b}, this regime is characterized by
equal accretion of big and small bodies, and can lead to order-unity
efficiency.  However, the mass spectrum below $\rst$ is no longer
frozen, and hence if this regime is reached it can wipe out the mass
spectrum laid down during trans-hill growth.

Trans-hill growth can also be terminated when {big bodies become
  separated by a Hill radius, in which case} oligarchy sets in.
Another complication {is collisional fragmentation}.  We leave more
detailed consideration of {these two effects} to future work.  { On
  the other hand, the complication of} gas damping is easily
incorporated by using an effective $s$.
 
Our theory explains results from previous studies of the asteroid and
Kuiper belt.  Most simulations to date adopt parameters relevant for
collisionless growth, typically with initial seeds $s \sim 1\km$. For
the Kuiper belt, this means the growth is collisionless at all times
and the efficiency is limited to $10\alpha \sim 0.1\%$. So if the
Kuiper belt has been severely dynamically depleted, it would initially
have had to contain multiple times the MMSN mass. Alternatively, the
efficiency of formation can approach unity if the initial seed size is
small. We explore how this may explain the formation of Kuiper belt in
an upcoming publication.

\acknowledgements

We thank Yanqin Wu for useful discussions.
We acknowledge   NSF grant
AST-1109776.

\bibliographystyle{apj}
\bibliography{kbo}

\appendix

\section{A. Numerical Solution of the Equations of motion}

In order to numerically integrate the equations of motion (Equations
(\ref{eq:gr})--(\ref{eq:repl})), we first re-write the equations in
terms of the trans-hill radius $\rth\equiv u/\sqrt{G\rho\alpha}$
(Equation (\ref{eq:rth})) and the rescaled time, 
   \be \tau ={\sigma\Omega\over\rho\alpha}t \ .
\label{eq:taudef}
\ee
which has  units
  of length.
Equation (\ref{eq:gr}) becomes
\be
{dR\over d\tau} = 
   \cases{
     (R/\rth)^2, & if  $R<\rth$ \cr
      R/\rth, & if  $R>\rth$ \cr
      }
      \label{eq:rdot}  \ ,
\ee
and Equation (\ref{eq:ustir}), with the replacement of Equation (\ref{eq:repl}), becomes 
\be
{1\over \rth}{d\rth\over d\tau}=
   -{\alpha \over s}
-{1\over\alpha}
\left(
 \int_{R<\rth} {R^6\over \rth^4}{dN\over\eta}+
 \int_{R>\rth}{R^3\over\rth}{dN\over\eta}
\right) \label{eq:utstir2}
\ee
The continuity equation (Equation (\ref{eq:cont})) becomes
\be
{\partial_\tau {N}} + V{\partial_R N}=0  \ ,
\label{eq:contapp}
\ee
where $V\equiv {dR/d\tau}$ is given by equation (\ref{eq:rdot}).

Our numerical method is as follows.  We choose an initial value for $\rtran$ and an initial form for the function $N(>R)$ on a logarithmic grid in $R$, where the grid typically has 400 gridpoints per decade of $R$. 
 According to Equation (\ref{eq:contapp}), the value of $N$ remains constant 
 at any $R$ that satisfies Equation (\ref{eq:rdot}).
 Therefore, for each initial $N$ on the grid, we integrate its corresponding $R$ according to Equation (\ref{eq:rdot}).   We simultaneously evolve $\rtran$ according to Equation (\ref{eq:utstir2}). 

\section{B. Self-similar solutions}

 If one considers either the collisionless 
 or  the collisional case, 
  then there is
 only a single characteristic radius, $\rth$. Therefore, Equation (\ref{eq:contapp}) may be written
 in self-similar form   when appropriately scaled.   We define the scaled independent variable to be the ratio of $R$ to its characteristic value,
\be
x\equiv {R/ \rth(\tau)} \ ,
\ee
and  the scaled dependent variable to be
\be
y(x)\equiv \rth^p N(R,\tau)/\eta \ ,
\ee
with $p$ to be determined below.
In scaled variables,  Equation (\ref{eq:contapp}) becomes
\be
{d\rth\over d\tau}\left( x{dy\over dx} +py\right)=V(x){dy\over dx} \ ,
\label{eq:dudtss}
\ee where \be V(x)= \cases{ x^2, & if $x<1$\cr x, & if $x>1$\cr } \ .
\ee Self-similarity demands \be {d\rth\over d\tau}=A \Rightarrow
\rth=A\tau \ee where $A$ is a constant to be determined.  Substituting
into Equation (\ref{eq:dudtss}), we may solve for $y$ to obtain \be
y(x) = y(1) \cases{ x^{-p}(A-x)^p/(A-1)^p, & $x<1$\cr x^{-pA/(A-1)}, &
  $x>1$ \cr } \ , \ee where we must have $A>1$.  Equation
(\ref{eq:utstir2}) becomes \be {A\over \rth}=-{\alpha\over
  s}-{1\over\alpha}\rth^{2-p} \left( \int_{x<1}x^6dy + \int_{x>1}x^3dy
\right) \ . \label{eq:uss} \ee It remains to determine the three
constants $p,A,y(1)$.  To do so, we consider the collisionless and
collisional cases separately.  

 In the collisionless case, $\alpha/s\ll
A/\rth$. This is equivalent to the condition $t \ll {\rho s/(\sigma \Omega)}$, 
i.e., that the time elapsed is less than the collision time between small bodies.
       Equation
(\ref{eq:uss})  demands that $p=3$  since $A$ is a
  constant in time, i.e., $N/\eta={y(x)/ \rth^3} $.  Note
that since $N$ is frozen (independent of time) at small $x$ (i.e. for
$R\ll \rth$), this immediately implies that $N\propto R^{-3}$ for
$R\ll \rth$.  Integrating Equation (\ref{eq:uss}) yields a relation
between the remaining two constants, $y(1)$ and $A$.  We write the
resulting self-similar solution as follows: \beqn
{N(>R,t)\over\eta}={\alpha\over\rth^3} g_{\rm nocol}(\gamma)
f_3\left({R\over\rth},\gamma\right) ,\ {\rm collisionless},
\label{eq:ss1}
\eeqn
where
\beqn
\rth(t)&=&{\sigma\Omega\over\rho\alpha}{t\over 1-3/\gamma} \\
g_{\rm nocol}(\gamma)&\equiv&{10\gamma^3\over \gamma^3+6\gamma^2+27\gamma+108}
\label{eq:Gam}
\\
f_p(x,\gamma)&\equiv&
 \cases{
x^{-p}\left(1-x(1-{p\over\gamma})\right)^p, & $x<1$\cr
x^{-\gamma}\left( {p\over\gamma} \right)^p, &   $x>1$ \cr
} \ . \label{eq:sslast} 
\eeqn

In the collisional case ($\alpha/s\gg A/\rth$), Equation
(\ref{eq:uss}) implies that $p=2$, which yields $N\propto R^{-2}$ for
$R\ll \rth$.  Integrating Equation (\ref{eq:uss}) yields the
self-similar solution \beqn
{N(>R,t)\over\eta}={1\over\rth^2}{\alpha^2\over s}
g_{\rm col}(\gamma) f_2\left({R\over\rth},\gamma\right) ,\ {\rm
  collisional}\
\label{eq:ncol}
 \eeqn
 where
\beqn
 \rth&=&{\sigma\Omega\over\rho\alpha}{t\over 1-2/\gamma}  \ ,
 \label{eq:rtcol} \\
g_{\rm col}(\gamma)&\equiv& {10\gamma(\gamma-3)\over \gamma^2+5\gamma+16} 
\label{eq:sscollast}
 \eeqn
 and $f_p(x,\gamma)$ is by Equation (\ref{eq:sslast}).

\end{document}